\documentclass[aps,12pt]{revtex4}
\usepackage[english]{babel}
\input epsf

\def\h{\hbar}

\begin{document}

%
%\begin{titlepage}
\title{Origin of the anomalies: The modified Heisenberg equation}
\author{J. G. Esteve }
\affiliation{
 Departamento de F\'{\i}sica Te\'orica,
 Facultad de Ciencias,
 Universidad de
 Zaragoza. Zaragoza, Spain. }
%
%
%
%\date{}
%
%\maketitle
%
\begin{abstract}
The origin of the anomalies is analyzed. It is shown that they are due
 to the fact that the generators of the symmetry do not leave invariant
 the domain of definition of the Hamiltonian and then a term,
 normally forgotten in the Heisenberg equation, gives an extra contribution
 responsible for the non conservation of the charges. This explanation
 is equivalent to that of the Fujikawa in the path integral 
 formalism. Finally, this formalism is applied to the conformal symmetry
 breaking in two-dimensional quantum mechanics.
\end{abstract}

\maketitle
%%
%\end{titlepage}

%\section{Introduction}
%%
The use of the symmetries of a system is one
 of the most fruitful techniques in physics, 
specially for quantum systems. Among the 
consequences of symmetry in quantum physics are: 
selection rules, relations between matrix 
elements of observables, degeneracies in 
energy and, specially, the existence of 
conservation laws which are guarantied by 
the Noether's theorem or by its equivalent, 
the Ehrenfest equation. In particular, the evolution
 of the expectation values of an operator $B$ 
is given by the Heisenberg equation
\begin{eqnarray}
{d \over dt} <\Psi(t)|B \Psi(t)> 
= &<\Psi(t)|{\partial B\over \partial t}
 \Psi(t)>
%\nonumber \\ & %
 + {i\over \h} <\Psi(t)|[H,B] \Psi(t)>\,,
\label{1}
\end{eqnarray}
that says that for any group of symmetry whose
 elements (or generators in the associated
Lie algebra) commute with the Hamiltonian 
$H$, if such operators do not depend explicitly
 on time $t$, their expectation value on any physical
 state must be constant with $t$.

Even for the case when the symmetry is not exact 
(it is explicitly broken), the Heisenberg equation (\ref{1}) 
gives us how do evolve with time the expectation 
values of the corresponding generators. This has 
been largely used, for example, in particle physics
 where the flavor symmetry is explicitly broken by
 mass terms, or in nuclear physics where isospin
 symmetry is broken by the Coulomb interaction between
 protons and by the up-down quark mass difference. 

However, there are some cases where although the 
symmetry is exact at the classical level, it is 
not preserved in the corresponding quantum theory. 
This is the anomalous symmetry breaking that was 
first discovered in quantum field theory when studding 
certain Feynman diagrams, and further in the 
analysis of $\pi^0 \rightarrow 2\gamma$ decay \cite{ABJ}, 
also in the Schwinger model \cite{Sch}. In the path 
integral formalism, the existence of 
anomalies can be viewed  as
 a consequence of that, in this case, even if 
the classical Lagrangian 
is invariant under the symmetry, the measure is 
not \cite{FUY}, that gives rise to extra surface 
terms which originate the anomaly.

In the Hamiltonian formalism, the anomaly can be understood
 as a consequence from the fact that the Heisenberg equation 
(\ref{1}) is exact only if the domain of definition 
of the Hamiltonian is invariant by the operator $B$; 
in other cases, it appears an extra term which is 
the responsible for the anomaly \cite{JGE}. To
 be more precise, let $H$ be the quantum Hamiltonian
 which is self-adjoint when defined on a domain $D_H$,
 then for any physical state in the Hilbert 
space ${\cal H}$ and any operator $B$ we have
\begin{eqnarray*}
{d \over dt} <\Psi(t)|B \Psi(t)> 
&=&<\Psi(t)|{\partial B\over \partial t}\Psi(t)>
+{i\over \hbar}	\left( <H\Psi(t)|B \Psi(t)> - 
<\Psi(t)|B H \Psi(t)>\right)
\end{eqnarray*}
which can be written as
\begin{eqnarray}
{d \over dt} <\Psi(t)|B \Psi(t)>=
<\Psi(t)|{\partial B\over \partial t}\Psi(t)> 
+ {i\over \h} 
<\Psi(t)|[H,B] \Psi(t)>+ {\cal A}\,,
\label{2}
\end{eqnarray}
where the anomaly ${\cal A}$ is defined as
\begin{eqnarray}
{\cal A}={i\over \h} <\Psi(t)|
\left( H^+ - H \right) B\Psi(t)>\,,
\label{3}
\end{eqnarray}
and in (\ref{2}) the commutator $[H,B]$ is defined 
in the whole Hilbert space. Equivalently, in the Heisenberg 
picture, we obtain that for any operator $B$ their derivative
 with respect to the time obeys a generalized Heisenberg equation:
\begin{eqnarray}
{d B\over d t}={\partial B\over \partial t} +
{i\over \h} [H,B] +{i\over \h} (H^+ -H)B\,.
\label{3b}
\end{eqnarray}

Comparing equations (\ref{1}) and (\ref{2}) we see 
that the Heisenberg equation (\ref{1}) is exact only whenever 
$B$ keeps invariant the domain of definition $D_H$ 
of the Hamiltonian, because if
$B\Psi_h \in D_H$ for  $\Psi_h \in D_H$, then the extra term 
$<\Psi_h(t)|\left( H^+-H \right) B\Psi_h(t)>=0$ 
as long as $H^+=H$ when acting 
on states of $D_H$. But when $B$ does 
not keep $D_H$ invariant, that is $B\Psi_h \notin D_H$,
the extra term will give a surface contribution responsible 
for the anomaly. In general, it is said that in
 the presence of an anomaly, the commutator of the Hamiltonian
 with the corresponding charges has two contributions, 
the regular one and an extra part originated by the anomaly,
 $[H,B]_{total}= [H,B]_{reg}+[H,B]_{anom}$. 
We see that the so called regular part is nothing but
 the commutator of the extension of the operators to the whole
 Hilbert space and the anomalous term is just $(H^+-H)B$.

It can be proved that the above description 
of the anomalies is equivalent to that of the path
 integral \cite{FUY}. For example, in quantum mechanics, 
the Feynman propagator is
\begin{eqnarray}
K(z,t;y,0)=\int_{\stackrel{x(0)=y}{ x(t)=z}} [d\mu(x)]
\exp{(i S(x,\dot{x})/ \h)}  
= \sum \!\!\!\!\!\!\!\int_n 
\varphi_n^*(y)\varphi_n(z) e^{-(iE_nt/ \h)}\,,
\label{4}
\end{eqnarray}
where $\varphi_n$ are the eigenvectors of the Hamiltonian
 and $\sum_n \!\!\!\!\!\!\!\!\int$ \, means sum 
over the discrete and integral over the continuum
 spectrum. In this sense, 
different self-adjoint extensions
 $H^{(\lambda)}$ of the quantum 
Hamiltonian associated to the same classical
 Lagrangian, give rise to different sets of
 orthonormal eigenvalues $\varphi_n^{(\lambda)}(x)$,
depending on the self-adjoint extension defined on
$D_H^{(\lambda)}$, 
and each of them is characterized by a different measure in 
the path integral version, so if a particular domain of
 definition $D_H^{(\lambda)}$
 is not invariant under the operator $B$, the same is true
 for the associated measure. The proof for quantum field 
theories is equivalent.

It should be noted that the existence of an anomaly
 is independent of the need or not of a regularization process
for the theory, as can be seen in some quantum mechanical
systems as that of a charged particle moving on a two-torus
 and coupled to an electromagnetic field, see \cite{NSM} and
 also \cite{AAE}.

Recently, there has been a renewed interest in 
anomalies in conformal quantum mechanics: the 3-dimensional
 ${1/ r^2}$ potential which is relevant as an example
 of an anomaly in Molecular Physics \cite{CGC},
 and the two dimensional $\delta$ interaction \cite{RJ}. In
 what follows, we shall analyze the later on the light of the
 generalized Heisenberg equation (\ref{2}). The problem is that 
of a free particle in two-dimensional quantum mechanics with a 
$\delta^2 (r)$ interaction.
\begin{eqnarray}
H={P^2 \over 2 m} +\lambda {1\over r}\delta(r)\,.
\label{5}
\end{eqnarray}
It can be seen that considering the extension of $H$
 to the whole Hilbert space $ {\cal H}=
 L^2(R_+,r dr)\bigotimes L^2(S_1,d\varphi)$, the 
theory is scale invariant, that is, the dilation 
operator $D= t H -G$ (where $G= ({1/ 4}) 
({\bf rp} +{\bf pr})$), the conformal generator $K=
 -t^2 H + 2 t D +({1/ 2}) r^2$ and $H$ close
 on commutation:
\begin{eqnarray}
{i \over \h} [K,D]&=& K\,,\\
{i \over \h} [H,K]&=& -2D\,,\\
{i \over \h} [D,H]&=&H\,,
\label{7}
\end{eqnarray}
showing that the invariance algebra is $SO(2,1)$. The above
equations together with the classical Heisenberg equation
 (\ref{1}) mean that $ {d\over dt} <K>={d\over dt}<D>=0$,
 and that there can not be any normalizable bound state.
 However, in order to properly define the quantum theory, 
we must first define the domain of definition $D_H$ of 
the Hamiltonian in such a way that ${\overline D_H} =\cal H$,
and that when acting on $D_H$ we have $H^+=H$. In order to do that, 
we start by removing the origin to avoid the singularities 
of the Hamiltonian, then working on 
${\bf \dot R^2} = {\bf R^2 / \{0,0\} }$ \cite{AGH}
and in polar coordinates, so the 
Hamiltonian reads:
\begin{eqnarray}
H=-{\h^2 \over 2m} \left( {\partial ^2 \over \partial r^2} + 
{1\over r} {\partial \over \partial r} + {1\over r^2} 
{\partial ^2 \over \partial \varphi^2} \right)\,.
\label{8}
\end{eqnarray}

The first question is to define the domain of 
definition of the operator ${d ^2 / d \varphi^2}$
which has deficiency indices $d_+=d_-=2$, so there are infinitely many 
self-adjoint extensions associated to different physical situations
. Here we shall start with the particular one 
with periodic boundary conditions,
 so defining   ${d ^2 /d \varphi^2}$ on
\begin{eqnarray}
d_0=\left\{ f(\varphi)\in L^2 (S_1,d\varphi) | f(0)=f(2\pi); 
f^\prime(0)=f^\prime(2\pi) \right \}\,,
\label{9}
\end{eqnarray}
in that case, acting on $d_0$ the eigenfunctions of
   ${d ^2 / d \varphi^2}$ are $\xi_n(\varphi)
= (2 \pi)^{-{1/ 2}}e^{i n \varphi}$ with $n\in Z$,
and then $D_H$ can be written as
\begin{eqnarray}
D_H= \bigoplus_{n\in Z}	\left( D_n(R_+,r dr)
\bigotimes \xi_n(\varphi)\right)\,,
\label {10}
\end{eqnarray}
where $D_n(R_+,rdr)$ must be chosen in such a way
 that the radial part of the Hamiltonian
\begin{eqnarray}
H_r=-{\h^2 \over 2 m} \left( {d^2\over dr^2} + {1\over r} {d\over dr}
 - {n^2 \over r^2}\right)\,,
\label{11}
\end{eqnarray}
is self-adjoint. For $n\neq 0$ the deficiency indices of $H_r$
 are (0,0)
 so it  is essentially self-adjoint acting on 
functions which vanishes at $r=0$ and infinity. But 
 for $n=0$, the deficiency indices are $(1,1)$ and 
there are infinitely many self-adjoint extensions
 characterized by a parameter $\beta$:
\begin{eqnarray}
D_{n=0}^\beta=\left\{ f(r)\in L^2(R_+,rdr)|
\lim_{r\rightarrow 0} \left({f(r)\over\log (\alpha_0 r)}\right)
=\beta\lim_{r\rightarrow 0}\left[ f(r)-\lim_{r^\prime \rightarrow 0}
\left({f(r^\prime)\over\log(\alpha_0 r^\prime)}\right)
\log(\alpha_0 r)\right]\right\},
\nonumber\\
\label{12}
\end{eqnarray}
where $	\alpha_0^2 =\left({2 m 
\Lambda_0/ \h^2 }\right)$, and $\Lambda_0$ is the 
dimensional constant we must introduce in order to make
 the deficiency index equations $H_r \Psi_\pm(r)=
 \pm i \Lambda_0 \Psi_\pm (r)$
dimensionally consistent. The case $\beta=0$ is
  the Friedrich's extension and corresponds to the 
situation $\lambda=0$ in (\ref{5}), whereas $\beta\neq 0$ accounts for the
case $\lambda\neq 0$. 

 Equation (\ref{12}) means that if
$f(r)\in D_{n=0}^\beta$ then for $r\rightarrow 0;\,\, f(r)\sim
a(\log(\alpha r)+b)+\theta(r)$ with
\begin{eqnarray}
{1\over\beta}= b +\log({\alpha\over\alpha_0})\,.
\label{14}
\end{eqnarray}
Now, it can be seen that for $n\neq 0$, we have $G\, D_n \subset D_n$
so $G$ leaves invariant $D_{n\neq 0}$, but for $n=0,\,\ \beta\neq 0$,
 and $f(r)\in D_0^\beta$, we obtain
\begin{eqnarray}
G\, f(r)\in D_0^{\beta^\prime} ;\,\,\,\,\, \beta^\prime={\beta\over
\beta+1}\,,
\label{15}
\end{eqnarray}
 so $D_0^\beta$ is not invariant by the action of $G$, and 
consequently it is not invariant by $D$ and $K$.
Hence the symmetry will be anomalously broken. 
The most relevant manifestation of this anomalous 
symmetry breakdown is the existence of a normalizable
 bound state
\begin{eqnarray}
\Psi_0(r,\varphi)={\alpha\over \pi^{1/2}} K_0(\alpha r)\,,
\label{16}
\end{eqnarray}
with energy
\begin{eqnarray}
E_0=-{\h^2\over 2m} \alpha^2\,,
\label{17}
\end{eqnarray}
where if $\Psi_0 \in D_0^\beta$ then $\alpha$ and 
$\beta$ are related by
${1/ \beta}=\log({\alpha / 2 \alpha_0}) 
+ \gamma$, ($\gamma $ is the Euler's constant). 
In this case
 for the dilation operator $D$, the left 
side of equation (\ref{2}) evaluates to $E_0$ 
whereas for the right side we have that
$<\Psi_0|{\partial D\over \partial t} \Psi_0>=
-{i\over\h}<\Psi_0|[H,D]\Psi_0>$, and  the
 equation (\ref{2}) reduces to
\begin{eqnarray}
{d\over d t} <\Psi_0|D\Psi_0>={\cal A}= {i\over \h}
<(H^+-H)D\Psi_0>\,.
\label{18}
\end{eqnarray}
Integrating by parts, we finally obtain
\begin{eqnarray}
{\cal A}&=& {\h ^2 \alpha ^2 \over 2 m}
\left\{ \left[r {d\over d r} K_0(\alpha r)
(r {d \over d r} +2)K_0(\alpha r)\right]_0^\infty
-  \left[ r  K_0(\alpha r){d\over d r}
(r {d \over d r} +2)K_0(\alpha r)\right]_0^\infty
\right\}\nonumber \\&=&-{\h ^2 \alpha ^2\over 2 m}\,.
\label{19}
\end{eqnarray}
Then we see that it is precisely the fact that $D$ 
does not keep $D_0^\beta$ invariant which originates 
the extra contribution which accounts for the value 
of the left side.

Finally, we can consider what happens with other self-adjoint 
extensions of the Hamiltonian. For example, defining 
the operator $ d^2 / d \varphi^2$ with vanishing boundary
 conditions on $d_v=\{f(\varphi)\in L^2(S_1,d \varphi) |
f(0)=f(2\pi)=0\}$, everything, respective to the anomaly, is
 equivalent to the case of periodic boundary conditions; but 
if we define $ d^2 / d \varphi^2$ on $d_\theta 
=\{f(\varphi)\in L^2(S_1,d \varphi) |
f(0)=e^{i 2\pi\theta} f(2\pi);
f^\prime(0)=e^{i 2\pi\theta} f^\prime(2\pi);\theta\in [0,1) \}$
 then the domain of
 definition of $H$ can be written as 
$D_H= \bigoplus_{n\in Z}(D_{n,\theta}(R_+,r dr)
\bigotimes (2\pi)^{-(1/2)}e^{i(n+\theta)\varphi})$, and for
$\theta\neq 0$ the deficiency indices of the radial part
 of the Hamiltonian
\begin{eqnarray}
H_r^\theta=-{\h^2\over 2m} \left( 
{d^2\over dr^2} + {1\over r}{d\over dr} -
{(n+\theta)^2\over r^2}\right)\,,
\label{20}
\end{eqnarray}
are $(0,0)$ for $n\neq 0,-1$, and $(1,1)$ for $n=0,-1$. In
this last case, 
defining the adequate domains $D_{0,\theta}^\beta$
 and  $D_{-1,\theta}^{\beta^\prime}$ it is easy to see that both
 subspaces are not invariant by $G$, which results in the fact
 that there are two normalizable bound states.

In conclusion, it has been shown that the origin of the anomalous
 symmetry breakdown is that the generators of the symmetry do not
 leave invariant the domain of definition of the Hamiltonian and then,
 although the formal commutator of those generators with $H$ vanishes,
 the charges are not conserved due to the extra surface
 term that appears in the exact Heisenberg equation (\ref{2}). 
For the case of the conformal symmetry breaking in the $\delta^2(r)$
potential, the anomaly has been calculated exactly and the 
Eq. (\ref{2}) verified. Similar results for the $1/r^2$
potential in three dimensional quantum mechanics will be discussed elsewhere.

{\bf Acknowledgments}

I would like to thank
M. Asorey,
for useful discussions.
This work has been supported the 
CICYT  grants  BFM2000-1057 
and FPA2000-1252.
%%

%%%%%%%%%%%%%%%%%%%%%%%%%%%%%%%%%%%%%%%%%%%%%%%%%%%%%%%%%%%%%%%%

%%%%%%%%%%%%%%%%%%%%%%%%%%%%%%%%%%%%%%%%%%%%%%%%%%%%%%%%%%%%%%%%
%%%%%%%%%%%%%%%%%%%%%%%%%%%%%%%%%%%%%%%%%%%%%%%%%%%%%%%%%%%%%%%%
%%
%%
%\newpage

%%

\end{document}